\documentstyle[aps,preprint]{revtex}
\tightenlines

\input psbox.tex
\begin{document}
\draft
\title{\bf Dirac Fermions on a Two-Dimensional Lattice
and the Intermediate Metallic Phase}
\author{K. Ziegler\\
Institut f\"ur Physik, Universit\"at Augsburg\\
D-86135 Augsburg, Germany\\
\today
%e-mail: ziegler@physik.uni-augsburg.de
}
\maketitle
\begin{abstract}
Consequences of different discretizations of the two-dimensional
Dirac operator on low energy properties (e.g., the number of nodes)
and their relations to gauge properties are discussed.
Breaking of the gauge invariance was suggested in a recent work by
M. Bocquet, D. Serban, and M.R. Zirnbauer [cond-mat/9910480]
in order to destroy an intermediate metallic phase of lattice
Dirac fermions with random mass. It is explained that such
a procedure is inconsistent with the underlying lattice physics.
Previous results point out that the logarithmic growth of the slope of
the average density of states with the system
size, obtained in the field-theoretical calculation of M. Bocquet
et al., could be a precursor for 
the appearence of an intermediate metallic phase.
\end{abstract}
%\pacs{PACS numbers: }

\section{Introduction}
\noindent
Two-dimensional (2D) Dirac (Majorana) fermions can be derived in statistical
physics (2D Ising model \cite{dotsenko}) and in models of condensed matter
systems (quasiparticles in $d$-wave superconductors \cite{hatsugai,nersesyan},
in the resonant valence bond state of the two-dimensional Heisenberg model
\cite{pepin}, or in the quantum Hall effect
\cite{semenoff,haldane,hatsugai1,ludwig}) always from lattice models.
The relation of all these lattice models and 2D Dirac fermions is based on
the agreement of their low-energy properties on certain length scales.
Using continuous Dirac fermions, which have a linear dispersion, their
$k=0$ node is identified with one of the nodes of the original lattice
model. However, there are technical reasons to study a discrete
(lattice) version of the Dirac operator because of problems related to the
unrestricted spectrum of the continuum model. In the presence of randomness
there is also the particular problem of correlations on a characteristic
length scale which restricts the use of the unrealistic white noise
randomness. 
In the renormalization group approach \cite{dotsenko} or in the saddle-point
approximation \cite{ziegler2} it is sufficient to specify the lattice by a
cutoff of the wavevector, assuming that only the $k=0$ node is important.
For a more detailed discussion of the model one must
define the lattice structure explicitly. The simplest case would be a nearest
neighbor (NN) discretization of the lattice difference operator
\begin{equation}
\nabla_jf(r)={1\over2}[f(r+e_j)-f(r-e_j)],
\label{nabla}
\end{equation}
where $e_j$ is the lattice unit vector in direction $j$ on a square lattice.
This choice, which was used in several papers by the author, exhibits an
``intermediate phase'', characterized by a non-vanishing density of states
(DOS), in the presence of a random mass for $|\langle m\rangle |\le m_c$.
It turned out that this phase is metallic with finite conductivity
\cite{ziegler4}.

The discretization (\ref{nabla}) was criticized in a
recent work by Bocquet et al. \cite{bocquet} because of the associated
sublattice symmetry. It was argued that the 
non-vanishing average DOS of this model \cite{ziegler4} is a sole consequence
of the fact that it belongs to a special universality class, and that a 
different discretization would have a vanishing average DOS.
Their argument is based on the renormalization group calculation 
of Ref. \cite{dotsenko} which gives a {\it linearly vanishing}
DOS at $E=0$.
(However, problems with this calculation exist because the RG has a
strange behavior in which the slope of the DOS {\it diverges}
logarithmically with the system size \cite{bocquet}.)
Since the non-vanishing DOS represents an intermediate metallic phase,
observed and discussed for several models in Refs. 
\cite{ziegler2,ziegler4,kagalovsky,senthil/marston,senthil,read},
the change of the discretization would result in a destruction of this
phase.

In the following, an interpretation of the 2D Dirac fermions with
discretization (\ref{nabla}) is given in terms of a tight-binding model
with flux $\pi$. It will be shown that the additional sublattice symmetry is
a special gauge symmetry. The destruction of the gauge symmetry by an
additional next-nearest neighbor term in (\ref{nabla}) leads to a
lattice model with an inconsistent gauge field representation.
More general, a change of the discretization affects physical
quantities like the Hall conductivity which depends on the number of nodes.
Finally, it will be shown, using a result from a previous calculation, that
breaking of the sublattice symmetry does not neccessarily result in a
vanishing average DOS at $E=0$. This indicates that even though
the properties of the intermediate metallic phase vary with a change of
the discretization, its very existence is rather stable.

\section{Discussion of the Model}

\noindent
The 2D Dirac Hamiltonian is given as
\[
H=H_0+m\sigma_3,\ \ H_0=
i\nabla_1\sigma_1
+i\nabla_2\sigma_2
\]
with Pauli matrices $\sigma_j$ and a random mass with
$\langle m_r\rangle=0$. $\nabla_j$ can be any discrete
lattice difference operator. It must be antisymmetric in order to have
a Hermitean matrix $H$. For instance, it could be a NN difference operator
(\ref{nabla}) or a combination of a NN  and a NNN difference operator.
The former will be called ``$\pi$ flux discretization'' and the latter
``flux breaking (FB) discretization'' for reasons which are explained
subsequently. The additional NNN difference operator adds new nodes to the
dispersion of the pure system. 
To understand the difference of the discretizations one can consider the
pure model, i.e. the Hamiltonian without the random mass.
Then the dispersion for the $\pi$ flux discretization is
\[
E_\pi=\pm\sqrt{\sin^2 k_1+\sin^2 k_2}
\]
which has nodes at $k_j=0,\pm\pi$. For the FB discretization
the dispersion is
\[
E_{FB}=\pm\sqrt{[\sin k_1+\sin (2k_1)]^2+[\sin k_2+\sin (2k_2)]^2}.
\]
with nodes at $k_j=0,\pm\pi,\arccos(-1/2)$.
Additional nodes of the pure sytem can increase the DOS of the disordered
system. This gives rise to the question whether all nodes contribute to the
DOS at low energy or only some. On the other hand, the nodes with non-zero
$k$ can be removed by adding a diagonal
hopping term to the Hamiltonian. As it was shown in a previous study 
\cite{ziegler3} (and will be discussed in Sect. III), the $k=0$ node
is sufficient for a non-zero average DOS at $E=0$. Thus the states with
wavevector $k\approx 0$ appear with non-vanishing density.

Two sublattices can be constructed by considering two layers of the original
square lattice, where the layers refer to the Dirac index $\alpha=1,2$. Then
the sublattices are defined through the condition that for the site
$(r,\alpha)$ either $r_1+r_2+\alpha$
is odd  (sublattice $\Lambda_1$) or even (sublattice $\Lambda_2$).
The Hamiltonian with $\pi$ flux discretization acts on these two sublattices
separately:
\begin{equation}
H=P_1HP_1+P_2HP_2,
\label{projection}
\end{equation}
where $P_j$ is the projector on $\Lambda_j$. Adding next-nearest neighbor
(NNN) terms to (\ref{nabla}) results in the Hamiltonian $H_{FB}$ which couples
both sublattices.

The transformation $H\to H S\sigma_3$ can be applied to the Hamiltonian with 
\[
S_{rr'}=(-1)^{r_1+r_2}\delta_{rr'}.
\]
$H$ commutes with $S\sigma_3$ \cite{ziegler3,bocquet}.
It changes the sign of $P_2HP_2$ but leaves $P_1HP_1$ unchanged.
This means it is a gauge transformation on $\Lambda_2$.

It is important to notice that property (\ref{projection}) does not mean that
one can project on one of these sublattices and ignore the other
projection because both sublattices are still statistically coupled
through the same random mass. This can be seen in the product of the
Green's function with its complex conjugate
$|(H+i\epsilon)^{-1}_{r\alpha,r'\alpha'}|^2$. It was shown \cite{ziegler4}
that for $\alpha'=\alpha+1\ (mod\ 2)$ the symmetry of $H$ implies
\[
|(H+i\epsilon)^{-1}_{r\alpha,r'\alpha}|^2
=-(H+i\epsilon)^{-1}_{r\alpha,r'\alpha}
(H+i\epsilon)^{-1}_{r\alpha',r'\alpha'}
\]
and
\[
|(H+i\epsilon)^{-1}_{r\alpha,r'\alpha'}|^2
=-(H+i\epsilon)^{-1}_{r\alpha,r'\alpha'}
(H+i\epsilon)^{-1}_{r\alpha',r'\alpha}.
\]
Thus $|(H+i\epsilon)^{-1}_{r\alpha,r'\alpha'}|^2$ on one sublattice
is the product of Green's functions from {\it both} sublattices.
The non-trivial properties of the two-particle Green's function are
a consequence of the Green's functions on the two sublattices which share
a common random mass.

The Hamiltonian $H_j=P_jHP_j$ describes a  tight-binding model on the
sublattice $\Lambda_j$ with hopping elements
\[
H_{r1,r\pm e_12}=H_{r2,r\pm e_11}=\pm i
\]
\[
H_{r1,r\pm e_22}=H_{r2,r\mp e_21}=\pm 1
\]
which represents a Hamiltonian with flux $\pi$.
Adding NNN terms to $\nabla_j$ changes this flux
because the new terms have the same phase factors
as the NN terms but on links which are twice as large as those of the NN
hopping terms. This leads to an inconsistency in the model because particles
would experience a magnetic field whose strength depends on the length of the
hops. Moreover, the invariance property $[H,S\sigma_3]_-=0$, which is the
gauge invariance on sublattice $\Lambda_2$, is violated by the NNN hopping
term.

It is possible to add NNN hopping terms to $H$ which are
consistent with the node structure of the $\pi$ flux discretization.
The simplest is
\[
h_{nnn}=H_0^2=(i\nabla_1\sigma_1+i\nabla_2\sigma_2)^2.
\]
It preserves the gauge field of the NN term.
Another one is 
\[
h_{nnn}'=H_0^2\sigma_3=(i\nabla_1\sigma_1+i\nabla_2\sigma_2)^2\sigma_3
\]
which does not preserve the gauge field because it contributes a positive
hopping element for $\alpha=1$ but a negative hopping element for $\alpha=2$.
Nevertheless, both NNN terms obey the condition
\begin{equation}
[h_{nnn},S\sigma_3]_-=[h_{nnn}',S\sigma_3]_-=0,
\label{gaugeinv}
\end{equation}
but only $h_{nnn}'$ satisfies the defining symmetry of class D of Altland
and Zirnbauer \cite{altland}
\[
h_{nnn}'=-\sigma_1{h_{nnn}'}^T\sigma_1.
\]
Property (\ref{gaugeinv}) can be used to estimate the non-vanishing
average DOS at $E=0$, as demonstrated in Refs. \cite{ziegler1,ziegler3}.

%%%%%%%%%%%%%%%%%%%%%%%%%%%%%% Random Walk %%%%%%%%%%%%%%%%%%%%%%%%%%%%%
\noindent
The discussion of the DOS  requires only the diagonal part of the 
one-particle Green's function 
$G_{rr,aa}=(H\pm i\epsilon\sigma_0)^{-1}_{rr,aa}$, which can be represented by
a sum over closed random walks \cite{glimm} beginning at $r$ and returning to 
it with the same Dirac index. Formally, the random walk representation can be
obtained from the hopping expansion of the Green's function
\[
(H_0+m\sigma_3+i\epsilon\sigma_0)^{-1}=(m\sigma_3+i\epsilon\sigma_0)^{-1}
\sum_{l\ge0}\Big[H_0(m\sigma_3+i\epsilon\sigma_0)^{-1}\Big]^l.
\]
This expansion is convergent for sufficiently large
$\epsilon$. After averaging over the random mass one can perform an analytic 
continuation to arbitrarily small $\epsilon>0$. 
The loops experience an effective gauge field because of complex hopping 
elements. An example is a simple plaquette (Fig. 1). In general
the flux per plaquette is $\phi=\pi$, in units of the flux quantum $\phi_0$.
An important property of the random walks follows from
\[
G(-m,i\epsilon)=\sigma_2G^T(m,i\epsilon)\sigma_2
\]
because this means that the random walks are reversed by the change
of the sign of the mass. An interpretation is that the currents in the
model can be reversed by reversing the sign of the mass, which is related
to the change of the sign of the Hall conductivity $\sigma_{xy}$ with $m$
\cite{semenoff}. 

%%%%%%%%%%%%%%%%%%%%%%%%%%%%%%%%%%%%%%%%%%%%%%%%%%%%%%%%%%%%%%%%%%%
\subsection{Hall Conductivity}
\noindent
A quantity which is sensitive to the type of discretization but robust
against disorder is the Hall conductivity $\sigma_{xy}$. It can be measured
as the linear response to an external gauge field \cite{semenoff}.
In a pure system ($m=const.$) with Hamiltonian
$H=m\sigma_3+h_1\sigma_1+h_2\sigma_2$ it reads \cite{ludwig}
\[
\sigma_{xy}={m\over2}
\int{1\over(m^2+h_1^2+h_2^2)^{3/2}}{d^2k\over(2\pi)^2}
\]
in units of $e^2/\hbar$. For $m\sim 0$ only the nodes contribute
significantly to the intergral:
\[
\sigma_{xy}(m)={m\over |m|}f(|m|),
\]
where $f(0)\ne0$ is proportional to the number of nodes. In the case
of the $\pi$ flux Dirac operator this gives just a Hall step for {\it each}
sublattice which means a single Hall step for the corresponding
$\pi$ flux tight-binding model. 
For the modified Hamiltonian $H_0+m\sigma_3+h_{nnn}'$ the Hall conductivity
is
\[
{1\over2}\int{m+\sin^2k_1+\sin^2 k_2
\over[(m+\sin^2k_1+\sin^2 k_2)^2+\sin^2k_1+\sin^2 k_2]^{3/2}}
{d^2k\over(2\pi)^2}
\] 
which gives
\[
\sigma_{xy}\sim const.+{1\over2\pi}{m\over |m|}\ \ \ (m\sim0).
\]
Thus, the additional term $h_{nnn}'$ contributes a constant Hall 
conductivity at small values of $m$. These results indicate that
approximations of the original model should not change the
number of nodes nor violate the structure of the Hamiltonian by
adding new terms of the type $h_{nnn}'$ in order to get the correct
Hall conductivity.

A general rule is that all nodes of the lattice model should be included
in the calculation of the low energy properties. This has severe consequence
for the renormaliztion group (RG) calculation which usually deals only with one
large length scale. The various nodes (different length scales) will create 
additional couplings under the RG transformation, leading eventually to 
a strong coupling behavior.
\\

%%%%%%%%%%%%%%%%%%%%%%%%%%%%%%%%%%%%%%%%%%%%%%%%%%%%%%%%%%%%%%%%%%%
\section{Model with single node}
\noindent
To study the contribution of the $k=0$ node to the average DOS,
all nodes of the $\pi$ flux discretization must be removed except for that
at $k=0$. This can be achieved by adding a diagonal term to $H$
\[
H_1=H+(\Delta-2)\sigma_3
\]
with $2\Delta f(r)=f(r+e_1)+f(r-e_1)+f(r+e_2)+f(r-e_2)$.
Obviously, this Hamiltonian
violates the sublattice gauge-invariance condition:
\[
\Delta\sigma_3S=-\sigma_3S\Delta.
\]
Without randomness (i.e. $m=0$) the dispersion is
\[
E_1(k_1,k_2)=\pm\sqrt{(\cos k_1+\cos k_2-2)^2+\sin^2k_1+\sin^2k_2}
\]
which is non-zero except for the node at $k_j=0$.
Now this Hamiltonian can be coupled by a random field $m'$ to the
Hamiltonian with nodes at $\pm\pi$
\[
H_2=H-(\Delta+2)\sigma_3
\]
as
\[
{\hat H}=\pmatrix{
H_1 & m'\sigma_3 \cr
m'\sigma_3 & H_2 \cr
}.
\]
Both Hamiltonians $H_1$ and $H_2$ violate the sublattice gauge-invariance
condition (\ref{gaugeinv}) but preserve the symmetry of class D
\[
{\hat H}=-\sigma_1{\hat H}^T\sigma_1.
\]
The Green's function can be projected on the subspace of $H_1$
\[
[H_1+i\epsilon-m'\sigma_3(H_2+i\epsilon)^{-1}\sigma_3m']^{-1}
\]
It was shown that this expression leads to a non-vanishing average DOS at
$E=0$ \cite{ziegler3}, indicating that the projection on one node
produces already a non-vanishing average DOS. 

\noindent
As a remark it should be noted that the pure 2D Ising model at the critical
point is  governed by the Hamiltonian \cite{ziegler2}
\[
H_{2DIM}=
a(1+a^2+2a\cos k_1-2\cos k_2)\sigma_3+2a(\sin k_1\sigma_1+\sin k_2\sigma_2)
\]
with $a=\sqrt2 -1$. This Hamiltonian is similar to $H_1$
as it has only one node at $k=0$.
\\

%%%%%%%%%%%%%%%%%%%%%%%%%%%%%%%%%%%%%%%%%%%%%%%%%%%%%%%%%%%%%%%%%%%%
\section{Conclusions}

The previous discussion is based on the one-particle Green's function
without referring to an effective field theory. This approach is motivated
by the aim to avoid additional symmetries which appear by the introduction
of new fields and which are not related to the original Hamiltonian or the
one-particle Green's function. In the case of the Hamiltonian $H$,
regardless of the discretization of $\nabla_j$, there are two
{\it discrete} symmetries:
\begin{equation}
H\to -\sigma_3 H \sigma_3
\label{symm1}
\end{equation}
for the ensemble and, therefore, for the {\it average} one-particle
Green's function. Moreover, there is a symmetry under the discrete
transformation
\begin{equation}
H\to -\sigma_1 H^T \sigma_1,
\label{symm2}
\end{equation}
which defines class D of Ref. \cite{altland}. It holds for each realization
of the model with Hamiltonian $H$. $h_{nnn}$ and $h_{nnn}'$, both break
symmetry (\ref{symm1}) whereas only $h_{nnn}$ breaks symmetry (\ref{symm2}).

From the symmetry point of view alone it is not entirely clear under which
conditions a vanishing average DOS at $E=0$ exist. The competition of
different nodes (i.e. different length scales) cannot be described only
by global symmetries but requires more detailed knowledge. 
However, there is a simple argument in terms of the hopping expansion which 
indicates that the number of {\it independent} random terms in the Dirac
Hamiltonian $H$ is crucial for the behavior of the average DOS around $E=0$.
Taking the zero-dimensional limit of $H$, i.e. the leading order of the
hopping expansion, there is a power law
\[
\langle \rho(E)\rangle\sim \rho_0|E|^\alpha,
\]
where $\alpha=0$ (random mass) and $\alpha=1$ (two-component random vector 
potential). The latter, of course, violates also the symmetry (\ref{symm2}).

The non-vanishing DOS for the class D model in $d=2$ with a diffusive
behavior was recently discussed by Senthil and Fisher \cite{senthil}
and Read and Green \cite{read} in terms of the RG flow of a non-linear
sigma model. Although this approach
was criticized by Bocquet et al. \cite{bocquet}, its result is in
agreement with that obtained with a different approach \cite{ziegler4}.
The appearence of the intermediate phase is quite natural for the 2D Dirac
fermions, since the pure model has a singular ``metallic'' phase at
$E=m=0$ with conductivity $e^2/h\pi$. This phase is robust against
a random vector potential, where the value of the conductivity remains
unchanged \cite{ludwig}. A random mass has apparently a stronger effect
because it reduces the conductivity at $m=0$ by a factor $1/(1+g/\pi)$,
where $g$ is the variance of the random mass \cite{ziegler4}. Moreover, it
broadens
the singular phase at $m=0$ to an interval $-m_c<m<m_c$ with non-vanishing
DOS. In terms of the random bond Ising model the non-vanishing DOS reflects
the existence of the Griffiths-McCoy-Wu phase, which cannot
be seen in the perturbative RG approach \cite{mccoy}.

The occurence of vortex-like excitations might be an important
effect, as suggested in Refs. \cite{bocquet,senthil,read}. 
Using the $\pi$ flux discretization, these vortices can be created 
by local edge currents in areas where the sign of the random mass changes:
An area with a positive mass has a positive Hall conductivity whereas
the surrounding area with negative mass has a Hall
conductivity with opposite sign. The resulting edge currents can have 
a long-range interaction. An effective model for this behavior can be
found in terms of the $Q$ matrix field theory of Ref. \cite{ziegler4}, in
which the random mass is replaced by a matrix field. Details will be 
published separately. 

In conclusion, the change of the discretization has a strong effect on
the node structure of the Dirac Hamiltonian. The correct discretization is
determined by the effective gauge field which the Dirac fermions experience. 
In the case of Dirac fermions with random mass, however, the average DOS at
low energies is relatively robust against the change of the discretization.
In particular, the $k\approx 0$ modes have a substantial contribution
to the average DOS. This indicates that also the intermediate metallic 
phase of the Dirac fermions with random mass and $\pi$ flux discretization 
should be robust under a change of the discretization.
\\

\noindent
Acknowledgement

\noindent
The author would like to acknowledge a communication with M.R. Zirnbauer
about his work in Ref. \cite{bocquet}. He is grateful to D. Braak
for interesting discussions. This work was supported by the
Sonderforschungsbereich 484.

\begin{figure}
\begin{center}\mbox{\psbox{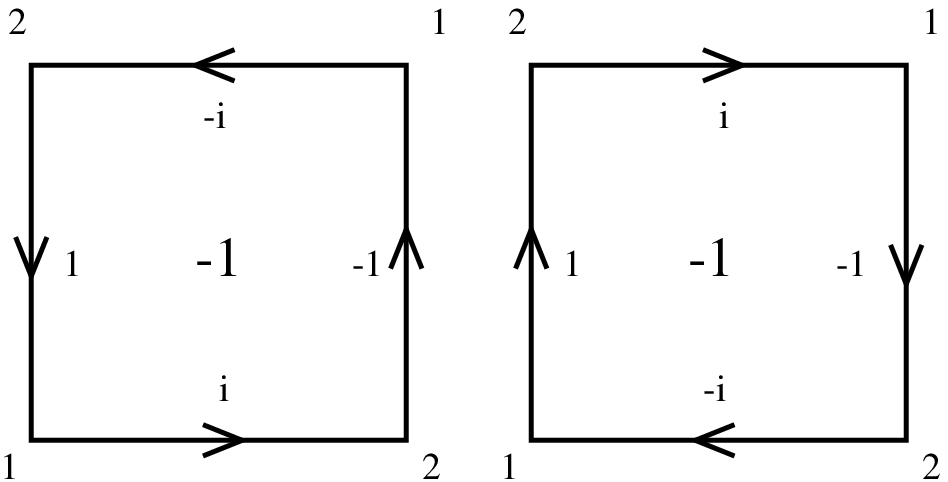}}\end{center}
\caption{Simple plaquettes are created from the Hamiltonian $H_0$ with
nearest neighbor discretization. They enclose a flux $\phi=\pi$ which
gives $e^{i\phi}=-1$. The sublattice gauge transformation 
$H\to HS\sigma_3$ does not change this property.}
\end{figure}

\end{document}